\documentclass[preprint,amsfonts,amssymb,amsmath,eqsecnum,%
tightenlines,nofootinbib,showpacs,showkeys]{revtex4}

\usepackage{amsthm}

\newcommand{\cN}{\mathcal{N}}

\newcommand{\cV}{\mathcal{V}}
\newcommand{\cW}{\mathcal{W}}
\newcommand{\bH}{\mathbf{H}}

\newcommand{\tH}{\tilde{H}}
 
\newcommand{\tP}{\tilde{P}}
\newcommand{\tcV}{\tilde{\cV}}

\newcommand{\Lsl}{\mathfrak{sl}} 

\newcommand{\Int}{\int\!\!}  
\newcommand{\ii}{\mathrm i}
\newcommand{\e}{\mathrm e}
\newcommand{\dd}{\mathrm d}
\newcommand{\sn}{\operatorname{sn}}
\newcommand{\cn}{\operatorname{cn}}
\newcommand{\dn}{\operatorname{dn}}

\begin{document}

\title{A New Family of $\cN$-fold Supersymmetry: Type B}
%REVTEX4
\author{Artemio Gonz\'{a}lez-L\'{o}pez} \email{artemio@fis.ucm.es}
\affiliation{Departamento de F\'{\i}sica Te\'{o}rica II,
  Facultad de Ciencias F\'{\i}sicas,\\
  Universidad Complutense, 28040 Madrid, Spain}
%\altaffiliation{Present address: }
\author{Toshiaki Tanaka} \email{totanaka@yukawa.kyoto-u.ac.jp}
\affiliation{Yukawa Institute for Theoretical Physics,\\
  Kyoto University, Kyoto 606-8502, Japan}
%\altaffiliation{Present address: }

\date{\today}

\begin{abstract}
  We construct a new family of $\cN$-fold supersymmetric systems which
  is referred to as ``type B''.  A higher derivative representation of
  the $\cN$-fold supercharge for this new family is given by a
  deformation of the type A $\cN$-fold supercharge.  By utilizing the
  same method as in the $\Lsl(2)$ construction of type A $\cN$-fold
  supersymmetry, we show that this family includes two of the
  quasi-solvable models of Post--Turbiner type.
\end{abstract}

\pacs{03.65.Ca; 03.65.Fd, 03.65.Ge; 11.30.Pb}%REVTEX4
\keywords{quantum mechanics; quasi-solvability; $\cN$-fold
  supersymmetry; intertwining relation}%REVTEX4

%\preprint{XXX-xxx}%REVTEX4

\maketitle%neELSART

\section{\label{sec:intro}Introduction}

Progress in the field of quasi-solvability in quantum systems (see
\cite{TuUs1,Sh89,Ushve,GKO94} and references therein) has recently
reached a new stage due to the discovery of the intimate relation
between quasi-solvability and $\cN$-fold supersymmetry.  An idea
essentially equivalent to $\cN$-fold supersymmetry was introduced for
the first time in Ref.~\cite{AnIoSp1} as an extension of ordinary
supersymmetric quantum mechanics~\cite{Witte1,Witte2}, and
investigated in various related
contexts~\cite{AnIoCaDe1,AnIoNi1,AnIoNi2,BaSa1,Samso1,BaSa2,Samso2,%
  Ferna1,Ortiz1,BaGaBhMi1,FeHu1,FeNeNi1,Plyus1,KlPl1}. The connection
of $\cN$-fold supersymmetry with quasi-solvability was first uncovered
in Ref.~\cite{AKOSW2} in an unexpected way, through the analysis of
the large order behavior of the perturbation series.  After similar
relations were found in several different
contexts~\cite{ASTY,KlPl2,KlPl3,DoDuTa1,DoDuTa2}, the equivalence
between quasi-solvability and $\cN$-fold supersymmetry was finally
proved in Ref.~\cite{AST2}.
 
Up to now, virtually all the $\cN$-fold supersymmetric models
explicitly constructed for \emph{arbitrary} $\cN$ belonged to the so
called \emph{type A} class, introduced in Ref.~\cite{AST1}.  From the
viewpoint of the connection with quasi-solvability, it was shown in
Refs.~\cite{ANST1,Tanak2} that type A $\cN$-fold supersymmetric systems are
essentially equivalent to the well-known quasi-solvable models
constructed from the $\Lsl(2)$ generators~\cite{Turbi1, GKO93, GKO94}.
Other recent developments in this respect can be found in
Refs.~\cite{SaTa1,ST1,CaIoNi1,AnSo1}.  In this Letter we construct a
new type of $\cN$-fold supersymmetric models which is a deformation of
(and hence different from) the type A class. We also show that the new
$\cN$-fold supersymmetric models are related to some of the
quasi-solvable models associated to the spaces of monomials classified
by Post and Turbiner~\cite{PoTu1}.
  
The article is organized as follows. In the next section we briefly
summarize $\cN$-fold supersymmetry and quasi-solvability. In
Section~\ref{sec:typeb} we introduce the type B $\cN$-fold supercharge
and construct an $\cN$-fold supersymmetric system with respect to it
by the same method used in the $\Lsl(2)$ construction of type A
$\cN$-fold supersymmetry~\cite{ANST1,Tanak2}. In
Section~\ref{sec:examp} several examples of the type B $\cN$-fold
supersymmetric models constructed in Section~\ref{sec:typeb} are
presented. The results obtained in this Letter and some open
problems they give rise to are discussed in the last section.

\section{\label{sec:defin}$\cN$-fold Supersymmetry}

First of all, we briefly review $\cN$-fold supersymmetry in
one-dimensional quantum mechanics. To this end, we introduce a bosonic
coordinate $q$ and fermionic coordinates $\psi$ and $\psi^{\dagger}$
satisfying
\begin{align}
  \{\psi,\psi\}=\{\psi^{\dagger},\psi^{\dagger}\}=0, \qquad
  \{\psi,\psi^{\dagger}\}=1.
%\label{eqn:}
\end{align}
The Hamiltonian $\bH_{\cN}$ is given by
\begin{align}
  \bH_{\cN}=H_{\cN}^{-}\psi\psi^{\dagger}
  +H_{\cN}^{+}\psi^{\dagger}\psi,
\label{eqn:NfHam}
\end{align}
where the components $H_{\cN}^{\pm}$ are ordinary scalar Hamiltonians,
\begin{align}
  H_{\cN}^{\pm}=\frac{1}{2}\,p^{2}+V_{\cN}^{\pm}(q),
\label{eqn:Ham+-}
\end{align}
with $p=-\ii\,\dd/\dd q$. $\cN$-fold supercharges $Q_{\cN}^{\pm}$ are
introduced by
\begin{align}
  Q_{\cN}^{-}=P_{\cN}^{-}\psi^{\dagger},\qquad
  Q_{\cN}^{+}=P_{\cN}^{+}\psi,
\label{eqn:dfNsc}
\end{align}
where the components $P_{\cN}^{\pm}$ are defined by
\begin{align}
  P_{\cN}^{-}=P_{\cN},\qquad P_{\cN}^{+}=(-1)^{\cN}P_{\cN}^{\,t}
\label{eqn:cmpsc}
\end{align}
in terms of an $\cN$th-order linear differential operator $P_{\cN}$ of
the form
\begin{align}
  P_{\cN}&=p^{\cN}-\ii\,w_{\cN-1}(q)\,p^{\cN-1}+\cdots+(-\ii)^{\cN-1}
  w_{1}(q)\,p+(-\ii)^{\cN}w_{0}(q)\notag\\
  &=(-\ii)^{\cN}\left(\frac{\dd^{\cN}}{\dd q^{\cN}}+w_{\cN-1}(q)
    \frac{\dd^{\cN-1}}{\dd q^{\cN-1}}+\cdots+w_{1}(q)\frac{\dd}{\dd q}
    +w_{0}(q)\right).
\label{eqn:Ndfop}
\end{align}
In Eq.~(\ref{eqn:cmpsc}), the superscript ${}^t$ denotes the
transposition of operators defined through a \textit{real} inner
product by $(A^{t}\phi,\psi)=(\phi,A\psi)$. For example, $p^{t}=-p$ on
a suitable space. Note that when all the functions $w_{k}$
appearing in Eq.~(\ref{eqn:Ndfop}) are real-valued, the operator $P_{\cN}^{+}$
defined by Eq.~(\ref{eqn:cmpsc}) is identical with the adjoint of
$P_{\cN}$: $P_{\cN}^{+}=P_{\cN}^{\dagger}$. Hence the above definition
is essentially equivalent to the ones in previous
articles~\cite{AST1,AST2,ANST1,Tanak2,ST1}.  The system
(\ref{eqn:NfHam}) is said to be \textit{$\cN$-fold supersymmetric} if
the following algebra holds:
\begin{gather}
  \{Q_{\cN}^{-},Q_{\cN}^{-}\}=\{Q_{\cN}^{+},Q_{\cN}^{+}\}=0,\\
  [Q_{\cN}^{-},\bH_{\cN}]=[Q_{\cN}^{+},\bH_{\cN}]=0.
%\label{eqn:}
\end{gather}
The former relation holds automatically due to the nilpotency of
$\psi$ and $\psi^{\dagger}$, while the latter is equivalent to the
following intertwining relations:
\begin{align}
  P_{\cN}^{-}H_{\cN}^{-}-H_{\cN}^{+}P_{\cN}^{-}=0, \qquad
  P_{\cN}^{+}H_{\cN}^{+}-H_{\cN}^{-}P_{\cN}^{+}=0.
\label{eqn:inter}
\end{align}
Therefore, the relations (\ref{eqn:inter}) give the condition for the
system $\bH_{\cN}$ to be $\cN$-fold supersymmetric. Note that the
Hamiltonians (\ref{eqn:Ham+-}) are always symmetric under the
transposition (on suitable spaces) even when they are not
hermitian, and thus each of the relations in Eq.~(\ref{eqn:inter})
actually implies the other. Note also that, due to the transposition
symmetry of the Hamiltonian, it was proved in Ref.~\cite{AnSo1} that
the anti-commutator of $Q_{\cN}^{-}$ and $Q_{\cN}^{+}$ defined by
Eqs.~(\ref{eqn:dfNsc}) and (\ref{eqn:cmpsc}) can be always expressed
as a polynomial of $\cN$th degree in $\bH_{\cN}$.

The $\cN$-fold supersymmetric models defined above have several
significant properties similar to those of the ordinary supersymmetric
models. One of the most notable ones is
quasi-solvability~\cite{AST2,ANST1,Tanak2}.  A differential operator
$H$ of a single variable $q$ is said to be \textit{quasi-solvable}
with respect to a given $\cN$th-order linear differential operator
$P_{\cN}$ of the form (\ref{eqn:Ndfop}) if it leaves invariant $\ker
P_{\cN}$, namely,
\begin{align}
  P_{\cN}H\cV_{\cN}=0,\qquad \cV_{\cN}=\ker P_{\cN}.
\label{eqn:qscon}
\end{align}
Then it can be easily shown~\cite{AST2} that an $\cN$-fold
supersymmetric system satisfying Eq.~(\ref{eqn:inter}) can always
be constructed from a quasi-solvable Hamiltonian $H$ by setting
$H_{\cN}^{-}=H$, $H_{\cN}^{+}=H+w'_{\cN-1}(q)$ and $P_{\cN}^{\pm}$ as
in Eq.~(\ref{eqn:cmpsc}). The converse is also true. Indeed, from the
intertwining relation (\ref{eqn:inter}) we find that all the
$\cN$-fold supersymmetric systems are quasi-solvable, the
quasi-solvability condition (\ref{eqn:qscon}) holding respectively for
$H=H_{\cN}^{\pm}$ and $P_{\cN}=P_{\cN}^{\pm}$.  When a system is
quasi-solvable but the elements of $\cV_{\cN}$ are not known
explicitly, the system is said to be \textit{weakly}
quasi-solvable~\cite{Tanak2}. More rigorous and sophisticated
definitions of quasi-solvability and related concepts can be found in
Ref.~\cite{Tanak3}.

\section{\label{sec:typeb}Type B $\cN$-fold Supersymmetry}

Recall \cite{AST1} that the type A $\cN$-fold supercharge is defined by an
$\cN$th-order linear differential operator $P_\cN$ of the form
\begin{align}
  P_{\cN}&=
  \prod_{k=-(\cN-1)/2}^{(\cN-1)/2}\bigl(p-\ii\,W
  +\ii\,k\,E\bigr)\notag\\
  &\equiv(-\ii)^{\cN}\left(\frac{\dd}{\dd q}+W-\frac{\cN-1}{2}
    E\right)\left(\frac{\dd}{\dd q}+W-\frac{\cN-3}{2}E
  \right)\times\cdots\notag\\
  &\quad\dots\times\left(\frac{\dd}{\dd q}+W
    +\frac{\cN-3}{2}E\right)\left(\frac{\dd}{\dd q}+W
    +\frac{\cN-1}{2}E\right),
\label{eqn:typeA}
\end{align}
where $W(q)$ and $E(q)$ are arbitrary (smooth) functions.  Consider next what
is perhaps the simplest deformation of the operator \eqref{eqn:typeA}, namely
\begin{align}
  P_{\cN}&=\left(p-\ii\,W+\ii\,F+\ii\,\frac{\cN-1}{2}E\right)
  \prod_{k=-(\cN-1)/2}^{(\cN-3)/2}\bigl(p-\ii\,W
  +\ii\,k\,E\bigr)\notag\\
  &=(-\ii)^{\cN}\left(\frac{\dd}{\dd q}+W-F-\frac{\cN-1}{2}
    E\right)\left(\frac{\dd}{\dd q}+W-\frac{\cN-3}{2}E
  \right)\times\cdots\notag\\
  &\quad\dots\times\left(\frac{\dd}{\dd q}+W
    +\frac{\cN-3}{2}E\right)\left(\frac{\dd}{\dd q}+W
    +\frac{\cN-1}{2}E\right).
\label{eqn:typeb}
\end{align}
The $\cN$th-order differential operator \eqref{eqn:typeb}, which depends on an
additional function $F(q)$, clearly reduces to the type A supercharge
\eqref{eqn:typeA} when $F$ vanishes identically. In this Letter we shall show
that if the functions $E$ and $F$ are related by the equation
\begin{align}
  F'(q)-E(q)F(q)+F(q)^{2}=0
\label{eqn:relEF}
\end{align}
the operator \eqref{eqn:typeb} defines a new type of $\cN$-fold supersymmetry.
Indeed, in this section we shall construct two Hamiltonians $H_{\cN}^{\pm}$
that are quasi-solvable with respect to the $\cN$-fold supercharges
$P_{\cN}^{\pm}$ defined by Eq.~(\ref{eqn:cmpsc}) in terms of the ``type B''
operator (\ref{eqn:typeb}).

To construct an $\cN$-fold supersymmetric model two different approaches can
be followed, namely the so called \textit{analytic} and \textit{algebraic}
methods~\cite{Tanak2}. We shall restrict ourselves in what follows to the
latter approach.  As in the case of type A $\cN$-fold
supersymmetry~\cite{ANST1,Tanak2}, to achieve our aim it is convenient to make
a suitable \textit{gauge} transformation and change of variable. Indeed, using
the following gauge potentials
\begin{align}
  \cW_{\cN}^{\pm}(q)=\frac{\cN-1}{2}\Int \dd q\,E(q)\mp\Int \dd q
  \,W(q),
\label{eqn:ggpot}
\end{align}
the type B $\cN$-fold supercharges $P_{\cN}^{\pm}$ are transformed
into
\begin{subequations}
\label{eqns:gtscs}
\begin{align}
  \tP_{\cN}^{-}&=\ii^{\cN}\,\e^{\cW_{\cN}^{-}}\,P_{\cN}^{-}\,
  \e^{-\cW_{\cN}^{-}} =\left(h'\right)^{\cN}\left(\frac{\dd}{\dd
      h}-\frac{1}{h}\right) \frac{\dd^{\cN-1}}{\dd h^{\cN-1}},
%\label{eqn:}
  \\
  \bar{P}_{\cN}^{+}&=\ii^{\cN}\,\e^{\cW_{\cN}^{+}}\,P_{\cN}^{+}\,
  \e^{-\cW_{\cN}^{+}}=\left(h'\right)^{\cN}\frac{\dd^{\cN-1}}{\dd
    h^{\cN-1}} \left(\frac{\dd}{\dd h}+\frac{1}{h}\right),
%\label{eqn:}
\end{align}
\end{subequations}
where $h(q)$ is a solution of the following differential equation:
\begin{align}
  h'(q)-F(q)h(q)=0.
\label{eqn:defFq}
\end{align}
As in Eqs.~(\ref{eqns:gtscs}), we shall hereafter attach tildes (bars)
to operators, vectors and vector spaces to indicate that they are
quantities gauge-transformed with the gauge potential $\cW_{\cN}^{-}$
($\cW_{\cN}^{+}$), respectively. From Eqs.~(\ref{eqn:relEF}) and
(\ref{eqn:defFq}), the relation between $h(q)$ and $E(q)$ reads
\begin{align}
  h''(q)-E(q)h'(q)=0,
\label{eqn:defEq}
\end{align}
which in turn is the same relation as the one for the type A case
employed in Refs.~\cite{AST1,ANST1,Tanak2}.  From
Eqs.~(\ref{eqns:gtscs}), we obtain gauge-transformed solvable
subspaces as
\begin{align}
  \tcV_{\cN}^{-}&=\ker\tP_{\cN}^{-}=\text{span\,}
  \left\{1,h,\dots,h^{\cN-2},h^{\cN}\right\},
\label{eqn:vspc-}
\\
\bar{\cV}_{\cN}^{+}&=\ker\bar{P}_{\cN}^{+}=\text{span\,}
\left\{h^{-1},h,h^{2},\dots,h^{\cN-1}\right\}.
\label{eqn:vspc+}
\end{align}
Let us begin with the construction of $\tH_{\cN}^{-}$. The operator
$\tH_{\cN}^{-}$ should be constructed so that it is quasi-solvable
with respect to $\tP_{\cN}^{-}$. This is achieved by finding a
second-order differential operator $\tH_{\cN}^{-}$ satisfying
\begin{align}
  \left(\frac{\dd}{\dd h}-\frac{1}{h}\right)\frac{\dd^{\cN-1}}{\dd
    h^{\cN-1}} \,\tH_{\cN}^{-}\,h^{k}=0\,, \qquad \forall
  k=0,1,\dots,\cN-2,\cN.
\label{eqn:qscd-}
\end{align}
For $\cN\geq 3$, there are six linearly independent differential
operators of order not greater than two solving
Eqs.~(\ref{eqn:qscd-}), namely:
\begin{subequations}
\label{eqns:qsop2-}
\begin{align}
  J_{--}&=\frac{\dd^{2}}{\dd h^{2}},\\
  J_{0-}&=h\frac{\dd^{2}}{\dd h^{2}}-(\cN-1)\frac{\dd}{\dd h},\\
  J_0 &=h\frac{\dd}{\dd h},\\
  J_{00}&=h^{2}\frac{\dd^{2}}{\dd h^{2}},\\
  J_{+0}&=h^{3}\frac{\dd^{2}}{\dd h^{2}}-(2\cN-3)h^{2}\frac{\dd}{\dd
    h}
  +\cN (\cN-2)h,\\
  J_{++}&=h^{4}\frac{\dd^{2}}{\dd h^{2}}-2(\cN-2)h^{3}\frac{\dd}{\dd
    h} +\cN (\cN-3)h^{2}.
\end{align}
\end{subequations}
Therefore, the general solution of (\ref{eqn:qscd-}) for $\cN\geq 3$
can be expressed as
\begin{align}
  \tH_{\cN}^{-}=-\sum_{\substack{i,j=+,0,-\\i\geq j}}a_{ij}^{(-)}
  J_{ij}+b_{0}^{(-)}J_{0}-C^{(-)},
\label{eqn:qsH-1}
\end{align}
where $a_{ij}^{(-)}$, $b_{0}^{(-)}$ and $C^{(-)}$ are constants.
Substituting Eqs.~(\ref{eqns:qsop2-}) into Eq.~(\ref{eqn:qsH-1}) we
obtain
\begin{align}
  \tH_{\cN}^{-}=-A_{4}^{-}(h)\frac{\dd^{2}}{\dd h^{2}}
  +A_{3}^{-}(h)\frac{\dd}{\dd h}-A_{2}^{-}(h),
\label{eqn:qsH-2}
\end{align}
with
\begin{subequations}
%\label{eqns:}
\begin{align}
  A_{4}^{-}(h)&=a_{++}^{(-)}h^{4}+a_{+0}^{(-)}h^{3}
  +a_{00}^{(-)}h^{2}+a_{0-}^{(-)}h+a_{--}^{(-)},\\
  A_{3}^{-}(h)&=2(\cN-2)a_{++}^{(-)}h^{3}+(2\cN-3)a_{+0}^{(-)}h^{2}
  +b_{0}^{(-)}h+(\cN-1)a_{0-}^{(-)},\\
  A_{2}^{-}(h)&=\cN(\cN-3)a_{++}^{(-)}h^{2}
  +\cN(\cN-2)a_{+0}^{(-)}h+C^{(-)}.
\label{eqn:A2-h}
\end{align}
\end{subequations}
On the other hand, the partner operator $\bar{H}_{\cN}^{+}$ should be
constructed so that it is quasi-solvable with respect to
$\bar{P}_{\cN}^{+}$. This is achieved by finding a second-order
differential operator $\bar{H}_{\cN}^{+}$ such that
\begin{align}
  \frac{\dd^{\cN-1}}{\dd h^{\cN-1}}\left(\frac{\dd}{\dd
      h}+\frac{1}{h}\right) \bar{H}_{\cN}^{+}\,h^{k}=0\,, \qquad
  \forall k=-1,1,2,\dots,\cN-1.
\label{eqn:qscd+}
\end{align}
For $\cN\geq 3$, Eqs.~(\ref{eqn:qscd+}) are solved by the following six
linearly independent differential operators of order less than or
equal to two:
\begin{subequations}
\label{eqns:qsop2+}
\begin{align}
  K_{--}&=\frac{\dd^{2}}{\dd h^{2}}-\frac{2}{h^{2}},\\
  K_{0-}&=h\frac{\dd^{2}}{\dd h^{2}}+\frac{\dd}{\dd h}-\frac{1}{h},\\
  K_0&=h\frac{\dd}{\dd h},\\
  K_{00}&=h^{2}\frac{\dd^{2}}{\dd h^{2}},\\
  K_{+0}&=h^{3}\frac{\dd^{2}}{\dd h^{2}}-(\cN-3)h^{2}\frac{\dd}{\dd h}
  -(\cN-1)h,\\
  K_{++}&=h^{4}\frac{\dd^{2}}{\dd h^{2}}-2(\cN-2)h^{3}\frac{\dd}{\dd
    h} +(\cN-1)(\cN-2)h^{2}.
\end{align}
\end{subequations}
Therefore, the general solution of (\ref{eqn:qscd+}) for $\cN\geq 3$
can be expressed as
\begin{align}
  \bar{H}_{\cN}^{+}=-\sum_{\substack{i,j=+,0,- \\ i\geq j}}a_{ij}^{(+)}
  K_{ij}+b_{0}^{(+)}K_{0}-C^{(+)},
\label{eqn:qsH+1}
\end{align}
where $a_{ij}^{(+)}$, $b_{0}^{(+)}$ and $C^{(+)}$ are constants.
Substituting Eqs.~(\ref{eqns:qsop2+}) into Eq.~(\ref{eqn:qsH+1}) we
have
\begin{align}
  \bar{H}_{\cN}^{+}=-A_{4}^{+}(h)\frac{\dd^{2}}{\dd h^{2}}
  +A_{3}^{+}(h)\frac{\dd}{\dd h}-A_{2}^{+}(h),
\label{eqn:qsH+2}
\end{align}
with
\begin{subequations}
%\label{eqns:}
\begin{align}
  A_{4}^{+}(h)&=a_{++}^{(+)}h^{4}+a_{+0}^{(+)}h^{3}
  +a_{00}^{(+)}h^{2}+a_{0-}^{(+)}h+a_{--}^{(+)},\\
  A_{3}^{+}(h)&=2(\cN-2)a_{++}^{(+)}h^{3}+(\cN-3)a_{+0}^{(+)}h^{2}
  +b_{0}^{(+)}h-a_{0-}^{(+)},\\
  A_{2}^{+}(h)&=(\cN-1)(\cN-2)a_{++}^{(+)}h^{2}-(\cN-1)a_{+0}^{(+)}h
  +C^{(+)}-\frac{a_{0-}^{(+)}}{h}-\frac{2a_{--}^{(+)}}{h^{2}}.
\label{eqn:A2+h}
\end{align}
\end{subequations}
If the operators (\ref{eqn:qsH-2}) and (\ref{eqn:qsH+2}) are
gauge-transformed back with the gauge potentials $\cW_{\cN}^{-}$ and
$\cW_{\cN}^{+}$, respectively, they are not in general Schr\"{o}dinger
operators of the form Eq.~(\ref{eqn:Ham+-}). The operators
$H_{\cN}^{\pm}$ assume the canonical form (\ref{eqn:Ham+-}) if and
only if the following conditions are fulfilled:
\begin{align}
  \frac{1}{2}\left(h'\right)^{2}&=A_{4}^{\pm}(h) \equiv
  P(h)=a_{4}h^{4}+a_{3}h^{3}+a_{2}h^{2}+a_{1}h+a_{0},
\label{eqn:cfcd1}
\\
A_{3}^{\pm}(h)&=\frac{\cN-2}{2}P'(h)\mp Wh'.
\label{eqn:cfcd2}
\end{align}
If the above conditions are satisfied, we have
\begin{align}
  H_{\cN}^{\pm}=\e^{-\cW_{\cN}^{\pm}}\,\bar{\tH}_{\cN}^{\pm}\,
  \e^{\cW_{\cN}^{\pm}}=-\frac{1}{2}\frac{\dd^{2}}{\dd
    q^{2}}+V_{\cN}^{\pm}(q),
%\label{eqn:}
\end{align}
where the potentials $V_{\cN}^{\pm}(q)$ are given by
\begin{align}
  V_{\cN}^{\pm}(q)=\frac{1}{2}\left[\left(\frac{\dd\cW_{\cN}^{\pm}(q)}{\dd
        q} \right)^{2}-\frac{\dd^{2}\cW_{\cN}^{\pm}(q)}{\dd
      q^{2}}\right] -A_{2}^{\pm}\bigl(h(q)\bigr).
\label{eqn:potfm}
\end{align}
From the second condition (\ref{eqn:cfcd2}) we obtain
\begin{align}
  -Wh'=-\frac{\cN}{2}a_{3}h^{2}+b_{1}h-\frac{\cN}{2}a_{1}\equiv Q(h),
\label{eqn:cfcd3}
\end{align}
where the constant $b_{1}$ is given by
\begin{align}
  b_{0}^{(\pm)}=(\cN-2)a_{2}\pm b_{1}.
\label{eqn:defb1}
\end{align}
The constants $a_{ij}^{(-)}$ and $b_{0}^{(-)}$ in $\tH_{\cN}^{-}$ are
related with the corresponding constants $a_{ij}^{(+)}$ and
$b_{0}^{(+)}$ in $\bar{H}_{\cN}^{+}$ by Eqs.~(\ref{eqn:cfcd1}) and
(\ref{eqn:defb1}).  To establish the relation between $C^{(-)}$ and
$C^{(+)}$, we will invoke the identity
$$
H_{\cN}^{+}-H_{\cN}^{-}=w'_{\cN-1}(q),
$$
where $w_{\cN-1}(q)$ is defined in Eq.~(\ref{eqn:Ndfop}). For the
$\cN$-fold supercharge of type B (\ref{eqn:typeb}) we have
$$
w_{\cN-1}(q)=\cN W(q)-F(q).
$$
Thus the condition for $H_{\cN}^{-}$ and $H_{\cN}^{+}$ to form an
$\cN$-fold supersymmetric pair reads
\begin{align}
  H_{\cN}^{+}-H_{\cN}^{-}=V_{\cN}^{+}(q)-V_{\cN}^{-}(q) =\cN
  W'(q)-F'(q).
\label{eqn:cdNfs}
\end{align}
On the other hand, from Eqs.~(\ref{eqn:ggpot}) and (\ref{eqn:potfm})
we have
\begin{align}
  V_{\cN}^{+}-V_{\cN}^{-}=W'-(\cN-1)EW-A_{2}^{+}(h)+A_{2}^{-}(h).
\label{eqn:dfpt1}
\end{align}
In order to rearrange the r.h.s.~of Eq.~(\ref{eqn:dfpt1}), the
following relations derived from Eqs.~(\ref{eqn:defFq}),
(\ref{eqn:defEq}), (\ref{eqn:cfcd1}) and (\ref{eqn:cfcd3}) are useful:
\begin{align}
  Q'(h)=-W'-EW,\qquad P'(h)=EFh.
\end{align}
With the aid of the above relations, we finally obtain
\begin{align}
  V_{\cN}^{+}-V_{\cN}^{-}=\cN W'-F'+(\cN-1)b_{1}-C^{(+)}+C^{(-)}.
%\label{eqn:}
\end{align}
Therefore, the condition for $\cN$-fold supersymmetry
(\ref{eqn:cdNfs}) holds when
\begin{align}
  C^{(+)}-C^{(-)}=(\cN-1)b_{1}.
\label{eqn:dfC+-}
\end{align}
In order to fix $C^{(\pm)}$, we write $A_{2}^{\pm}(h)$ as follows:
\begin{align}
  A_{2}^{\pm}(h)=A_{21}(h)\pm A_{22}(h).
%\label{eqn:}
\end{align}
From Eqs.~(\ref{eqn:A2-h}), (\ref{eqn:A2+h}) and (\ref{eqn:dfC+-}) we
have
\begin{subequations}
\label{eqns:A22A21}
\begin{align}
  A_{22}(h)&=\frac{P'(h)}{2h}-\frac{P(h)}{h^{2}}+\frac{\cN-1}{2}Q'(h),\\
  A_{21}(h)&=\frac{(\cN-1)(\cN-2)}{12}P''(h)-\frac{P(h)}{h^{2}}
  -\frac{Q(h)}{\cN h}+R,
\end{align}
\end{subequations}
where the constant $R$ is given by
\begin{align}
  R=-\frac{(\cN+1)(\cN-4)}{6}\,a_{2}+\frac{b_{1}}{\cN}
  +\frac{1}{2}\left(C^{(+)}+C^{(-)}\right).
\label{eqn:adC+-}
\end{align}
In this case, $C^{(\pm)}$ are determined from Eqs.~(\ref{eqn:dfC+-})
and (\ref{eqn:adC+-}) as
\begin{align}
  C^{(\pm)}=\frac{(\cN+1)(\cN-4)}{6}a_{2} \pm\frac{\cN^{2}-\cN\mp
    2}{2\cN}b_{1}+R.
%\label{eqn:}
\end{align}
Summarizing the results obtained so far, the gauge-transformed
operators of the type B $\cN$-fold supersymmetric Hamiltonians are given by
\begin{multline}
  \bar{\tH}_{\cN}^{\pm}=-P(h)\frac{\dd^{2}}{\dd h^{2}}+\left[
    \frac{\cN-2}{2}P'(h)\pm Q(h)\right]\frac{\dd}{\dd h}
  -\left\{\frac{(\cN-1)(\cN-2)}{12}P''(h)\right.\\
  -\frac{P(h)}{h^{2}}-\frac{Q(h)}{\cN h}
  \pm\left.\left[\frac{P'(h)}{2h}-\frac{P(h)}{h^{2}}
      +\frac{\cN-1}{2}Q'(h)\right]+R\right\}.
%\label{eqn:}
\end{multline}
The type B potentials $V_{\cN}^{\pm}$ are calculated by substituting
Eqs.~(\ref{eqn:ggpot}) and (\ref{eqns:A22A21}) into
Eq.~(\ref{eqn:potfm}). In terms of $h$ we have
\begin{align}
  V_{\cN}^{\pm}(h)=&-\frac{1}{12P(h)}\left\{(\cN^2-1)\left[P(h)P''(h)
      -\frac{3}{4}\bigl(P'(h)\bigr)^{2}\right]-3Q(h)^{2}\right\}
  +\frac{P(h)}{h^{2}}+\frac{Q(h)}{\cN h}\notag\\
  &{}\pm\left[\cN\frac{P'(h)Q(h)-2P(h)Q'(h)}{4P(h)}
    -\frac{P'(h)}{2h}+\frac{P(h)}{h^{2}}\right]-R,
%\label{eqn:}
\end{align}
while in terms of $q$
\begin{align}
  V_{\cN}^{\pm}(q)&=\frac{1}{2}W(q)^{2}-\frac{1}{\cN}F(q)W(q)
  +\frac{1}{2}F(q)^{2}-\frac{\cN^{2}-1}{24}
  \left[2E'(q)-E(q)^{2}\right]\notag\\
  &\quad{}\pm\frac{1}{2}\bigl[\cN W'(q)-F'(q)\bigr]-R.
\label{eqn:tBptq}
\end{align}
We note again that the potential (\ref{eqn:tBptq}) reduces to the type
A form~\cite{ST1,Tanak2} if we set $F(q)=0$.

\section{\label{sec:examp}Examples}

In this section we shall exhibit some examples of the
$\cN$-fold supersymmetric models of type B constructed in the previous section.
As the first example, we choose $P(h)=2(h-h_{0})$. In this case
$Q(h)=b_{1}h-\cN$ from Eq.~(\ref{eqn:cfcd3}). Using
Eqs.~(\ref{eqn:defFq}), (\ref{eqn:defEq}), (\ref{eqn:cfcd1}) and
(\ref{eqn:cfcd3}) we obtain
\begin{subequations}
%\label{eqns:}
\begin{gather}
  h(q)=q^{2}+h_{0},\quad E(q)=\frac{1}{q},\quad
  F(q)=\frac{2q}{q^{2}+h_{0}},\\
  W(q)=-\frac{b_{1}}{2}q+\frac{\cN-b_{1}h_{0}}{2q}.
\end{gather}
\end{subequations}
The pair of potentials (\ref{eqn:tBptq}) reads
\begin{align}
  V_{\cN}^{\pm}(q)&=\frac{b_{1}^{2}}{8}q^{2}+\frac{(\cN\mp\cN
    -b_{1}h_{0}-1)(\cN\mp\cN-b_{1}h_{0}+1)}{8q^{2}}\notag\\
  &\quad{}+(1\pm 1)\frac{q^{2}-h_{0}}{(q^{2}+h_{0})^{2}}
  -\frac{b_{1}}{4}(\cN\pm\cN -b_{1}h_{0})
  +\frac{b_{1}}{\cN}-R.
%\label{eqn:}
\end{align}

In the next example, we choose $P(h)=\nu^2(h-h_{0})^{2}/2$. In
this case, $Q(h)=b_{1}h+\cN\nu^2 h_{0}/2$ from
Eq.~(\ref{eqn:cfcd3}).  Employing again Eqs.~(\ref{eqn:defFq}),
(\ref{eqn:defEq}), (\ref{eqn:cfcd1}) and (\ref{eqn:cfcd3}) we obtain
\begin{subequations}
%\label{eqns:}
\begin{gather}
  h(q)=\e^{\nu q}+h_{0},\quad E(q)=\nu,\quad
  F(q)=\frac{\nu}{1+h_{0}\e^{-\nu q}},\\
  W(q)=-\frac{(2b_{1}+\cN\nu^2)h_{0}}{2\nu}\,\e^{-\nu q}
  -\frac{b_{1}}{\nu}.
\end{gather}
\end{subequations}
The pair of potentials (\ref{eqn:tBptq}) now reads
\begin{align}
  V_{\cN}^{\pm}(q)&=\frac{(2b_{1}+\cN\nu^2)^{2}h_{0}^{2}}{8\nu^2}\,
  \e^{-2\nu q}+\frac{(2b_{1}+\cN\nu^2)(2b_{1}\pm \cN\nu^2)h_{0}}{4\nu^2}\,
  \e^{-\nu q}\notag\\
  &\quad{}-(1\pm 1)\frac{\nu^2 h_{0}\e^{-\nu q}}{2\left(1+h_{0}
      \e^{-\nu
        q}\right)^{2}}+\frac{b_{1}^{2}}{2\nu^2}+\frac{b_{1}}{\cN}
  +\frac{\cN^{2}+11}{24}\nu^2 -R.
%\label{eqn:}
\end{align}

In our final example we take
$$
P(h)=\frac{\nu^2}2\,(1-h^2)(1-k^2\,h^2)\,,
$$
so that $Q(h)=b_1h$ and
\begin{subequations}
\begin{gather}
  h(q) = \sn(\nu q)\,,\quad E(q)= -\frac{\nu\sn(\nu
    q)\big(k'^2+2k^2\cn^2(\nu q)\big)}{\cn(\nu q)\dn(\nu q)}\,,\\
  F(q)=\frac{\nu\,\cn(\nu q)\dn(\nu q)}{\sn(\nu q)}\,,\quad
  W(q) = -\frac{b_1}{\nu}\,\frac{\sn(\nu q)}{\cn(\nu q)\dn(\nu q)}\,.
\end{gather}
\end{subequations}
In the latter formulas $\sn$, $\cn$, and $\dn$ denote the Jacobi
elliptic functions of modulus $k$ (with $0\leq k\leq 1$),
$k'=\sqrt{1-k^2}$ is the complementary modulus, and $\nu$ is a
positive constant. The pair of $\cN$-fold supersymmetric potentials
(\ref{eqn:tBptq}) constructed from this choice of $P(h)$ is
\begin{align}
 V_{\cN}^\pm(q)&=\frac12\,(1\mp1)\,\nu^2 k^2\sn^2(\nu q)
 +\frac12\,(1\pm1)\,\frac{\nu^2}{\sn^2(\nu q)}\notag\\
 &\quad{}+\frac{\big(2b_1-k'^2\nu^2(\pm\cN-1)\big)
   \big(2b_1-k'^2\nu^2(\pm\cN+1)\big)}{8\,\nu^2 k'^2\cn^2(\nu q)}\notag\\
 &\quad{}-\frac{\big(2b_1+k'^2\nu^2(\pm\cN-1)\big)
   \big(2b_1+k'^2\nu^2(\pm\cN+1)\big)}{8\,\nu^2 k'^2\dn^2(\nu q)}\notag\\
 &\quad{}+\frac{\nu^2}{12}(1+k^2)(\cN^2-7)+\frac{b_1}{\cN}
 \Big(1\pm\frac{\cN^2}2\Big)-R\,.
\end{align}

\section{\label{sec:discu}Discussion}

In this article, we have constructed a new family of $\cN$-fold
supersymmetry in which a higher-derivative representation of the
$\cN$-fold supercharge is given by Eq.~(\ref{eqn:typeb}) with the
constraint (\ref{eqn:relEF}). In view of quasi-solvability, it turns
out that the gauged Hamiltonians obtained here correspond to
quasi-solvable operators of the type investigated by Post and
Turbiner~\cite{PoTu1}. More precisely, $\tH_{\cN}^{-}$ in
Eq.~(\ref{eqn:qsH-1}) is identical with the case C operator in
\cite{PoTu1}, while $\bar{H}_{\cN}^{+}$ in Eq.~(\ref{eqn:qsH+1})
without $K_{--}$ is equivalent to the case D operator in \cite{PoTu1}.
The reason why the operator $K_{--}$ does not appear in
Ref.~\cite{PoTu1} is that the authors considered only differential
operators with polynomial coefficients. From our point of view,
however, it is evident that the operator $K_{--}$ is indispensable for
$\bar{H}_{\cN}^{+}$ to be the $\cN$-fold supersymmetric partner of
$\tH_{\cN}^{-}$. Indeed, without $K_{--}$ the number of independent
parameters in $\bar{H}_{\cN}^+$ differs from the one in $\tH_{\cN}^{-}$.
This is clearly impossible, since any differential operator
$\bar{H}(h)$ leaving the space (\ref{eqn:vspc+}) invariant is
equivalent to an operator $\tH(h)$ preserving the space
(\ref{eqn:vspc-}) under the transformation
$\bar{H}(h)=h^{\cN-1}\tH(h^{-1})h^{-\cN+1}$.  An interesting fact is
that the $\cN$-fold supersymmetric systems of type B characterized by the
$\cN$-fold supercharge (\ref{eqn:typeb}) and by the potentials
(\ref{eqn:tBptq}) connect, by the formal limit $F(q)\to 0$, the
quasi-solvable models of Post--Turbiner type~\cite{PoTu1} with the
$\Lsl(2)$ quasi-solvable models~\cite{Turbi1}, which are essentially
equivalent to the type A $\cN$-fold supersymmetric systems.

The algebraic construction with the constraint (\ref{eqn:relEF})
presented in this article turns out to be especially useful when the
solvable subspace can be gauge-transformed into a space of monomial
type.  With the use of this method it is possible to construct, as an
application, the most general $\cN$-fold supersymmetry whose solvable
sector is spanned by monomials~\cite{GoTa2}.

On the other hand, a direct calculation of the intertwining relation
(\ref{eqn:inter}) with the type B $\cN$-fold supercharge indicates
that the condition (\ref{eqn:relEF}) imposed in this article may be
only sufficient but not necessary for the existence of type B
$\cN$-fold supersymmetry. This suggests that the class of type B
potentials may be wider than the quasi-solvable models of
Post--Turbiner type obtained here. One of the reasons for this is that
the framework of $\cN$-fold supersymmetry makes sense even when the
solvable subspace has no known analytic expression, that is, when the
Hamiltonian is weakly quasi-solvable. Therefore, it may also be
possible that, once we have a system of $\cN$-fold supersymmetry
constructed from an $\cN$-dimensional vector space $\tcV_{\cN}$ with
the above procedure, it turns out that this system can be extended
in such a way that the solvable sector is no longer given by the starting
vector space $\tcV_{\cN}$. Work on these and related issues is in progress and
will be reported elsewhere.

\begin{acknowledgments}%REVTEX4
%\begin{ack}%ELSART
  This work was partially supported by the DGI under grant
  no.~BFM2002--02646 (A. G.-L.) and by a JSPS research fellowship (T.
  T.). One of us (T. T.) would like to thank A. A. Andrianov and M.
  Sato for useful discussions, as well as all the members
  of the Departamento de F\'{\i}sica Te\'{o}rica II, Universidad
  Complutense, for their kind hospitality during his stay.
\end{acknowledgments}%REVTEX4
%\end{ack}%ELSART

%\appendix

%\bibliography{nfsusy}

\end{document}